\documentstyle[aps,prl,floats,epsf,preprint,tighten]{revtex}

\begin{document}
\preprint{\vtop{\hbox{RU02-2-B}\hbox{MCTP-02-06}\hbox{hep-th/0202138}
\vskip24pt}}

\title{Angular Momentum Mixing in Crystalline Color Superconductivity}

\author{Ioannis Giannakis${}^{a}$,
James T.~Liu${}^{b}$
and Hai-cang Ren${}^{a}$
\footnote{E-mail: giannak@summit.rockefeller.edu, jimliu@umich.edu,
ren@summit.rockefeller.edu}}

\address{
${}^{(a)}$ Physics Department, The Rockefeller University\\
1230 York Avenue, New York, NY 10021-6399\\[4pt]
${}^{(b)}$ Michigan Center for Theoretical Physics\\
Randall Laboratory, Department of Physics, University of Michigan\\
Ann Arbor, MI 48109--1120}

\maketitle

\begin{abstract}
In crystalline color superconductivity, quark pairs form at non-zero
total momentum.  This crystalline order potentially enlarges the domain
of color superconductivity in cold dense quark matter.  We present a
perturbative calculation of the parameters governing the crystalline
phase and show that this is indeed the case.  Nevertheless, the
enhancement is modest, and to lowest order is independent of the
strength of the color interaction.
\end{abstract}


\ifpreprintsty\else
\begin{multicols}{2}
\fi
\narrowtext


While Quantum Chromodynamics (QCD) has now been firmly established as
the theory of strong interactions, many important questions, such as the
nature of confinement and chiral symmetry breaking, still remain.
Attempts to address such issues have led to recent interest in
exploring the nature of nuclear matter under extreme conditions of
temperature and density.
New phases of nuclear matter have long been sought after, both
theoretically and experimentally.  Under the right conditions,
nuclear matter is expected to undergo a phase
transition to a quark-gluon plasma.  At cold temperatures,
however, recent application of QCD suggests a much richer phase
structure of quark matter than originally anticipated.
Perhaps the most prominent of such phases, and one which may play an
important r\^ole in the core of neutron stars, is that of color
superconductivity \cite{barrois,arw}.

Color superconductivity is essentially the quark analog of
BCS superconductivity.  Because of attractive quark-quark interactions
in QCD, the Fermi sphere of quarks becomes unstable against the
formation of Cooper pairs and the system becomes superconducting at
sufficiently low temperatures.
Quantitative results may be obtained at ultra-high baryon densities
where asymptotic freedom ensures the validity of the weak coupling
approximation.  In this regime, the main contribution arises
from one-gluon exchange in the color antisymmetric channel.  The
resulting transition temperature, $T_c$, is given in terms of the
running QCD coupling $g(\mu)$ by
\cite{son,SW,Pisarski,bio,hong,ren,liu,nfl,wan}
\begin{equation}
k_BT_c=512\pi^3e^\gamma\Big({2\over N_f}\Big)^{5\over 2}
{\mu\over g^5}e^{-\sqrt{{6N_c\over N_c+1}}{\pi^2\over g}
-{\pi^2+4\over 16}(N_c-1)},
\label{eqvirus}
\end{equation}
where $\mu$ is the chemical potential and $\gamma=0.5772\ldots$ is
the Euler constant.

This result that $T_c\sim e^{-\kappa/g}$ is remarkable for its
non-BCS behavior (where BCS theory instead predicts $T_c^{({\rm BCS})}
\sim e^{-\lambda/g^2}$).  The relation between $T_c$ and the zero
temperature gap energy $\Delta_0$, on the other hand, remains BCS-like:
$\Delta_0=\pi e^{-\gamma}k_B T_c$. This non-BCS scaling is driven by
the poor screening of the long range color magnetic interaction. The
same mechanism is also responsible for non-Fermi liquid behavior
which suppresses $T_c$ significantly \cite{ren,liu,nfl}.

Since the main attractive channel is between different species of
quarks, search for a realistic color superconductivity phase must take
into consideration the masses of the light quarks.  This effect, not
accounted for in Eq.~(\ref{eqvirus}), is especially important when
considering the color flavor-locked state \cite{arw} involving $u$,
$d$ and $s$ quarks.  Although not strictly identical, this situation
is well modeled by separating the Fermi surfaces of massless quarks
that participate in the pairing.  This case was recently investigated
in Refs.~\cite{bowers,kundu,leb,ong}, where it was shown that a new
crystalline superconductivity state may arise for an appropriate window
of separation between Fermi surfaces.

Crystalline superconductivity was originally investigated
in BCS theory, where ferromagnetic impurities
separate the Fermi surfaces of each electron spin \cite{larkin,fulde,takada},
so that $\mu_{1,2}=\mu\pm\delta$.  When the system is cooled from the normal
phase, the transition temperature is the maximum pairing temperature,
allowing for a possible net momentum $2\vec q$ for a pair.  For
a sufficiently small separation, $\delta\le\delta_c$, zero momentum
pairing wins.  However, for $\delta$ in the range $\delta_c<\delta<
\delta_{\rm max}$, pairing with non-zero momentum is favored.  The net
momentum provides a region of phase space where both quarks may sit on
their respective Fermi surfaces.  However it breaks rotational
invariance and leads to crystalline order.  This state is also known as
the LOFF state, after the work of Refs.~\cite{larkin,fulde}.  Finally,
for $\delta>\delta_{\rm max}$, the transition temperature drops to zero,
and superconductivity is lost.

In the BCS case, both $\delta_c$ and $\delta_{\rm{max}}$ as well as
the pair momentum $q_{\rm{max}}$ (at $\delta=\delta_{\rm max}$) are
proportional to the initial ({\it i.e.}\ $\delta=0$) zero temperature
gap energy, with the constants
of proportionality being independent of the details of the interaction. 
In particular, $\delta_c^{({\rm BCS})}\approx0.606\,\Delta_0$, while
$\delta_{\rm max}^{({\rm BCS})}\approx0.754\,\Delta_0$.  As this indicates,
the window for BCS crystalline superconductivity is fairly narrow.

In this letter, we explore the crystalline superconductivity state
for color superconductivity.  Our approach follows the perturbative method
developed in Refs.~\cite{ren,liu} which allows a systematic approach to
determining $T_c$ (but not directly the zero-temperature gap).  We find
that for color superconductivity, while $\delta_c$ remains unchanged,
$\delta_{\rm max}$ is modestly enhanced, so that there is a widening of
the region of crystalline order compared to the BCS case.  As in
the behavior of $T_c$ itself in Eq.~(\ref{eqvirus}), the resulting
non-BCS behavior of the color LOFF state also traces its origin to
the long range nature of the magnetic gluons.


We begin with a brief review of the perturbative method developed in
Refs.~\cite{ren,liu} for a normal-phase determination of the transition
temperature, however
now applied to pairing with total momentum $2\vec q$ and shifted
chemical potentials $\mu\pm\delta$.  The proper vertex function for
quark pair scattering in the color antisymmetric channel is given by a
Dyson-Schwinger equation
\begin{eqnarray}
\Gamma_{{\vec q}, \delta}(P_f|P_i)=
\Gamma^{2PI}_{{\vec q}, \delta}(P_f|P_i)
\ifpreprintsty\else
\kern11em\nonumber\\
\fi
+{1\over {\beta}}\sum_{\nu}\int{{d^3{\vec p}}\over {(2\pi)^3}}
K_{{\vec q}, \delta}(P_f|P)\Gamma_{{\vec q}, \delta}(P|P_i),
\label{eqvioc}
\end{eqnarray}
where the initial and final momenta are ${\vec q}\pm{\vec p}_i$ and
${\vec q}\pm{\vec p}_f$. Here $P$ stands for Euclidean four-momentum,
$({\vec p}, -{\nu})$ with $\nu$ being the Matsubara energy,
and the four-momentum of the pair is $Q= (2{\vec q}, 0)$.
Eq.~(\ref{eqvioc}) is of the Fredholm type
with kernel
\begin{eqnarray}
K_{{\vec q}, \delta}(P^{\prime}|P)=\Gamma^{2PI}_{{\vec q}, \delta}
(P^{\prime}|P)
\ifpreprintsty\else
\kern11em\nonumber\\
\times
\fi
S(Q+P, {\mu}+{\delta})
S(Q-P, {\mu}-{\delta}),
\label{eqvuron}
\end{eqnarray}
where $\Gamma^{2PI}$ denotes the two quark irreducible part of the vertex
and $S(P, {\mu})$ the full quark propagator with chemical
potential $\mu$.

The pairing instability may be probed by examining the
spectrum of the kernel, given by the eigenvalue equation
\begin{equation}
Ef(P)=\sum_{\nu}\int{{d^3{\vec p}^{\prime}}\over {(2\pi)^3}}
K_{{\vec q}, \delta}(P|P^{\prime})f(P^{\prime}),
\label{eqzouros}
\end{equation}
for eigenfunctions $f(P)$.  At weak coupling, all eigenvalues are
less than one for sufficiently high temperature.  Then, as the
temperature is lowered, a point is reached where the largest eigenvalue
reaches unity, $E_{\rm max}(T,\mu,\delta,q)=1$, and the solution to
Eq.~(\ref{eqvioc}) becomes singular, corresponding to the onset of the
pairing instability.

We have indicated explicitly the dependence of $E_{\rm max}$ on the
temperature, mean chemical potential,
Fermi surface separation and the pair momentum. The condition $E_{\rm
max}=1$ may thus be
solved to obtain a pairing temperature as a function of $q$ for given
$\mu$ and $\delta$; this provides a determination of $T_c(\mu,\delta)$
when maximized over $q$.  Ordinary and crystalline order corresponds to
$q=0$ and $q\ne0$, respectively, at this maximum point.  At weak
coupling, the relevant $\delta$ and $q$ are both much smaller than the
mean chemical potential $\mu$, and the onset of LOFF order can be
treated as a perturbation from ordinary pairing at $\delta=q=0$. Thus we
look for an expansion of the eigenvalue $E_{\rm max}$ in ascending powers
of $g$ (with coefficients that may depend on $\ln g$). Through
Eq.~(\ref{eqvirus}), this is equivalent to an expansion in descending powers
of $\ln{\mu/(k_BT_c^0)}$, where $T_c^0$ is the $\delta=0$
value. Quantities such as ${{k_BT_c^0}/{\mu}}$,
${{\delta}/{\mu}}$ and ${q/{\mu}}$ are much smaller than any
finite power of $g$ and can be neglected.

To the accuracy of Eq.~(\ref{eqvirus}), the contribution of antiparticles to
the scattering amplitude can be ignored and the two quark irreducible
part of the vertex $\Gamma^{2PI}$ is well approximated by single
gluon exchange:
\begin{eqnarray}
{\Gamma}_{{\vec q}, \delta}(P^\prime|P)\simeq-{i\over {g^2}}
(1+{1\over N})\bigl[D_{M}({\vec k}, \omega)N_M({\vec p}, {\vec p}\,^{\prime},
{\vec k}\,)
\ifpreprintsty\else
\kern4pt\nonumber\\
\fi
+D_{E}({\vec k}, \omega)N_E({\vec p}, {\vec p}\,^{\prime}, {\vec
k}\,)\bigr],
\label{eqacioz}
\end{eqnarray}
where ${\vec k}={\vec p}\,^{\prime}-{\vec p}$ and $\omega=\nu^{\prime}-\nu$.
Here $N_M$ and $N_E$ are kinematic factors and $D_{M}({\vec k}, \omega)$
and $D_{E}({\vec k}, \omega)$ are magnetic and electric gluon propagators
with hard dense loop (HDL) resummation.  While the one-loop quark self
energy contributes to the prefactor of Eq.~(\ref{eqvirus}), it will not
interfere with the perturbative contribution of nonzero $\delta$ and $q$.
Consequently we replace $S$ with the free quark propagator,
$S({\nu}, p; {\mu})=i/(i\,{\nu}-p+{\mu})$, throughout.

The eigenvalue problem, Eq.~(\ref{eqzouros}), was previously analyzed
within each angular momentum channel for $\delta=q=0$ \cite{liu}. The
highest eigenvalue with angular momentum $J$ reads
\begin{equation}
E_J=1+{2\over {{\ln{1\over {\epsilon}}}}}\Bigl( 3c_J+{\ln{T\over T_c^0}}\Bigr)
+O\left(\ln^{-2}{\textstyle{1\over {\epsilon}}}\right),
\label{eqxion}
\end{equation}
where
\begin{equation}
c_J=\int_{-1}^{1}{dx}{{P_J(x)-1}\over {1-x}}
=\cases{0,&if $J=0$;\cr
        -2\sum_{j=1}^{J}{1\over j}, &if $J\ne 0$,\cr}
\label{eqtricky}
\end{equation}
and $\epsilon=k_BT_c^0/\hat\mu$ where $\hat\mu=256\pi^3
g^{-5}(2/N_f)^{5/2}\mu$ is a rescaled chemical potential.
The corresponding eigenfunction is
\begin{equation}
f_{JM}(P)={2\pi\over\sqrt{\ln{T_c^0\over\epsilon T}}}
\sin\left({\pi\ln{1\over\hat\nu}
\over 2\ln{T_c^0\over\epsilon T}}\right){Y_{JM}(\hat p)\over p},
\label{eqeigenfunc}
\end{equation}
where $\hat\nu=\nu/\hat\mu$
and $Y_{JM}(\hat p)$ are spherical harmonics. Since the level
spacing for different $J$ is of the same order as the perturbation to
be introduced, a degenerate perturbation method has to be employed.

Nonzero $\delta$ and $q$ modify the pole structure of the quark
pair propagator of Eq.~(\ref{eqvuron}) in a similar manner as in the BCS
case. In addition, a nonzero $\delta$ will shift the Debye mass in
the HDL resummed denominators of the vertex, and the numerators of the
one-gluon vertex, $N_M$ and $N_E$, depends on the total momentum
$\vec q$.  Both corrections to the vertex function are of the order of
$({{\delta}/ {\mu}})^2$ or $({{q}/ {\mu}})^2$ and may be ignored.

To diagonalize the kernel, Eq.~(\ref{eqvuron}), with nonzero $\delta$ and $q$
to $O({\ln^{-1}{1\over {\epsilon}}})$, we evaluate $K$ between
the $O(1)$ eigenfunction $f_{JM}(P)$ and its adjoint
\begin{equation}
{\overline f}_{JM}(P)=S(P)S(-P)
f_{JM}(P)\simeq{f_{JM}(P)\over{{\nu^2}+(p-\mu)^2}},
\label{eqriabov}
\end{equation}
and define
\begin{equation}
E_{J^{\prime}M^{\prime}, JM}
={1\over {{\beta}^2{\Omega}^2}}
\sum_{P,P^{\prime}}{\overline f}_{J^{\prime}
M^{\prime}}(P^{\prime})K_{{\vec q}, \delta}
(P^{\prime}|P)f_{JM}(P).
\label{eqfiou}
\end{equation}
Using the eigenvalue equation, Eq.~(\ref{eqzouros}), for $\delta=q=0$
and to leading order in $g$, we find
\begin{eqnarray}
E_{J^{\prime}M^{\prime}, JM}=
{\delta}_{JJ^{\prime}}{\delta}_{MM^{\prime}}
+{2\over {\ln{1\over\epsilon}}}
\langle J^{\prime}M^{\prime}|h_{op}|JM\rangle.
\label{eqtzwrtzo}
\end{eqnarray}
The operator $h_{op}$ is given by $3c_{op}+v_{op}$, with $c_{op}$
diagonal in the angular momentum representation, $\langle
J^{\prime}M^{\prime}|c_{op}|JM\rangle=3c_J{\delta}_{JJ^{\prime}}
{\delta}_{MM^{\prime}}$, and $v_{op}$ diagonal in the
coordinate (angle) representation,
\begin{equation}
v_{op}={\ln{T\over T_c^0}}+{\psi}\Bigl({1\over 2}\Bigr)
-{\rm Re}\,{\psi}\Bigl({1\over 2}-{i\over{2\pi}}{\beta}
(\delta-q\,{\cos\theta})\Bigr),
\label{eqgarbage}
\end{equation}
where $\psi(\zeta)={d\over {d{\zeta}}}{\ln{\Gamma(\zeta)}}$. Here $\theta$
is the polar angle with respect to momentum $\vec q$. The azimuthal
quantum number $M$ remains conserved.  Both $c_{op}$ and $v_{op}$ are
Hermitian and $c_{op}$ is non-positive.  For $q=0$, the rotation symmetry
is intact and the largest eigenvalue of $h_{op}$ still corresponds to
$J=0$. Here we recover the BCS result
\begin{equation}
E_{0}=1
+{2\over {{\ln{1\over {\epsilon}}}}}\Bigl[{\ln{{T_c^0}\over T}}
+{\psi}\Bigl({1\over 2}\Bigr)-
{\rm Re}\,{\psi}\Bigl({1\over 2}-{i\over {2\pi}}{\beta}
\delta\Bigr)\Bigr].
\label{eqtzwrtzopou}
\end{equation}

With ${\vec q}\ne 0$, different angular momentum components start to mix
and this widens the LOFF window with respect to $\delta$. In what follows,
we shall focus on the component $M=0$ since the additional sign change
introduced by nonzero $M$ will not be the most favorable. On returning
to a coordinate representation, $Y_{JM}(\hat p)\to
u(x\equiv\cos\theta)$, the eigenvalue equation for $h_{op}$ becomes
\begin{equation}
3\int_{-1}^1dx^\prime{u(x^\prime)-u(x)\over|x-x^\prime|}
+v_{op}(x)u(x)=\lambda u(x),
\label{eqint}
\end{equation}
where the first term on the left hand side is the coordinate
representation of $c_{op}$, following from the integral representation
of $c_J$ in Eq.~(\ref{eqtricky}).

To explore the upper threshold of the
LOFF window, $\delta_{\rm{max}}$, we take the limit $T\to 0$, whereupon
the function $v_{op}(x)$ reduces to
\begin{equation}
v_{op}(x)={\psi}\Bigl({1\over 2}\Bigr)
-\ln{\delta_{\rm{max}}\over 2\pi k_BT_c^0}-{\ln}|1-{x\over x_c}|,
\label{eqlimv}
\end{equation}
with $x_c=\delta/q$. The only possibility for obtaining a diverging
$\delta_{\rm{max}}$ is to take full
advantage of the logarithmic singularity
of $v_{op}$ as $x\to x_c$. Consider, {\it e.g.}, a trial function
\begin{equation}
u(x)=\cases{{1\over\sqrt{2\alpha}}, & for $x_c-\alpha<x<x_c+\alpha$;\cr
        0, &otherwise.\cr}
\label{eqtricky2}
\end{equation}
In the limit $\alpha\to 0$, we find $u^\dagger v_{op}
u\to\ln{1\over\alpha}$ but $u^\dagger c_{op}u\to-\ln{1\over\alpha}$,
so that $u^\dagger h_{op}u\to -2\ln{1\over\alpha}$. Therefore the
operator $h_{op}$ is bounded from above, and the widening of the LOFF
window cannot grow indefinitely.  A numerical analysis reveals
that the largest eigenvalue of the integral equation, Eq.~(\ref{eqint}),
with the limiting $v_{op}$ of Eq.~(\ref{eqlimv}) for a fixed $\delta$ is
$\lambda\simeq-\ln{\delta\over 2\pi k_BT_c^0}-1.304$, which occurs at
$q\simeq 1.16\,\delta$.  The corresponding $u(x)$ is shown
in Fig.~\ref{fig1}, demonstrating the peak at $x=x_c$.  For a second
order phase transition, the eigenvalue equation coincides with the limit
of the gap equation as $T\to T_c$.  An
anisotropy will develop for the order parameter in the superphase as well.
Applying the pairing condition, $E_{\rm max}=1$, and the relation between
$T_c^0$ and $\Delta_0$, we find
\begin{equation}
\delta_{\rm{max}}\simeq 1.67\,k_BT_c^0\simeq 0.96\,\Delta_0,
\label{eqresult}
\end{equation}
to be compared with the BCS case, $\delta_{\rm{max}}^{({\rm BCS})}
\simeq 0.754\,\Delta_0$.

\begin{figure}[t]
\ifpreprintsty
\epsfxsize 12cm
\else
\epsfxsize\hsize
\fi
\centerline{\epsffile{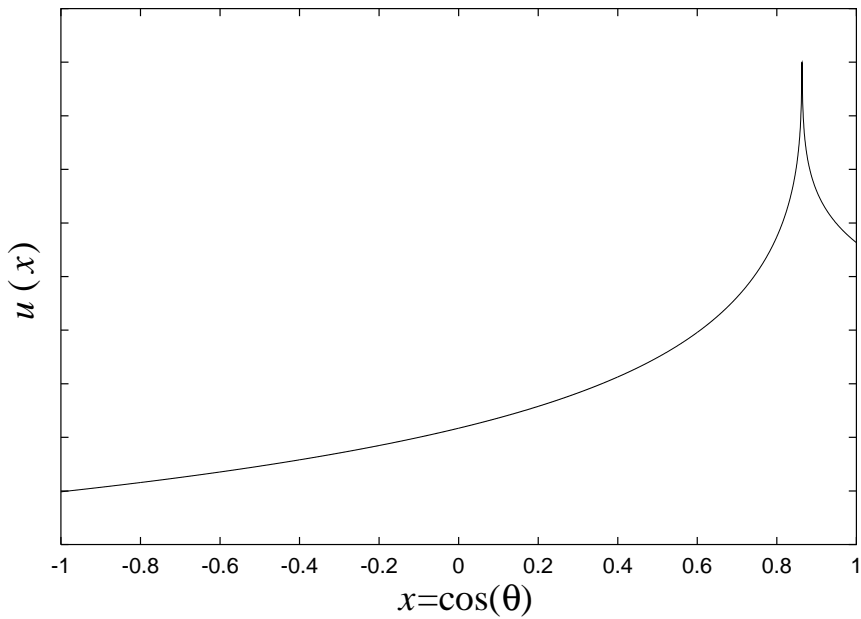}}
\medskip
\caption{The eigenfunction $u(x)$ corresponding to the largest eigenvalue
of Eq.~(\ref{eqint}).  For fixed $\delta$, this occurs at $x_c = 0.863$.}
\label{fig1}
\end{figure}

The determination of the onset, $\delta_{c}$, is numerically
more tedious, but an upper bound can be readily obtained. For very
small $\delta$, we have $v_{op}\simeq {\ln}{T\over {T_c^0}}
- {{7{\zeta}(3)}\over {2{\pi}^2}}{\beta}^2(\delta-q\,{\cos\theta})^2$.
The last term is too small to mix different $J$, and we have $J=0$
for pairing. Since $({{{\partial}E_0}\over {{\partial}q^2}})_{\delta,
\beta} < 0$ at $q=0$, the LOFF state is not favored. Then if we assume
the onset of the LOFF state starts from $q_{c}=0$ for sufficiently
large $\delta$, the critical value of $\delta$ follows from the condition
$({{{\partial}E_0}\over {{\partial}q^2}})_{\delta, \beta}=0$.
Applying Eq.~(\ref{eqtzwrtzopou}) and the pairing condition
$E_{\rm max}=1$, we obtain
\begin{equation}
\delta_c \simeq 0.606\,{\Delta}_0,
\label{eqrionc}
\end{equation}
which is identical to the BCS result.
Although we have not been able to verify it analytically, numerical
results indicate that pairing indeed starts at $q=0$.  The numerically
computed phase diagram is shown in Fig.~\ref{fig2}.

\begin{figure}[t]
\ifpreprintsty
\epsfxsize 12cm
\else
\epsfxsize\hsize
\fi
\centerline{\epsffile{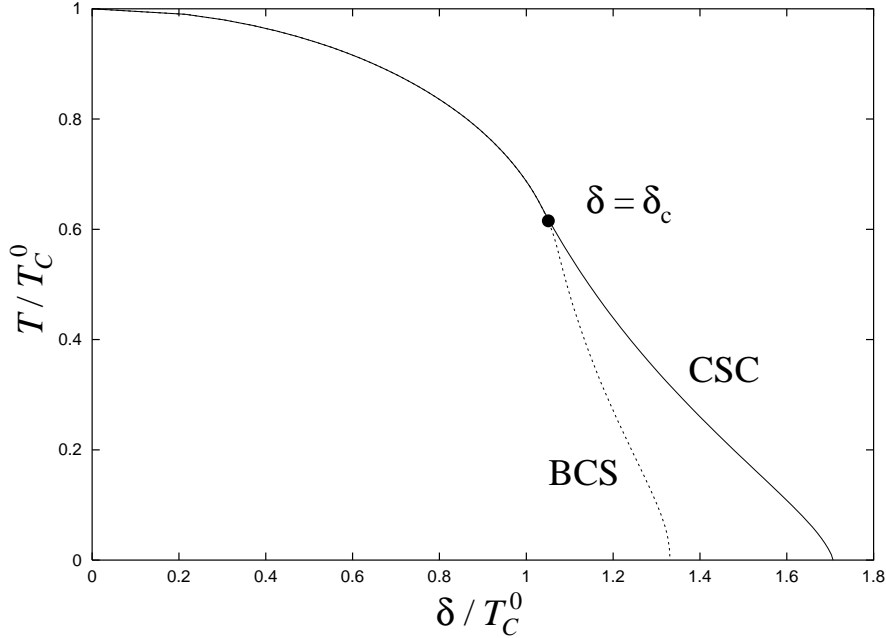}}
\medskip
\caption{The transition temperature, $T$, as a function of $\delta$.  The
BCS transition is shown for comparison.}
\label{fig2}
\end{figure}

We note that Ref.~\cite{leb} suggested that there would be an indefinite
widening of the window for the color LOFF state by drawing an analog
with a $1+1$ dimensional system on account of the forward singularity
of the scattering vertex.  The present results, however, demonstrate
that while the window is indeed widened over that of the BCS case,
the widening is fairly modest, as indicated in Fig.~\ref{fig2}.
This result may be understood since the reduction of the available
phase space for pairing with increasing $\delta$
in higher dimensions dominates over any enhancement arising from a
$1+1$ dimensional analog.  Furthermore, unlike the isotropic BCS case,
different angular momenta have to be considered simultaneously towards a
determination of $\delta_{\rm{max}}$, and the long range order parameter
itself would not be spherical. To our knowledge, such a mixing effect
is entirely new and the approximations we have used are well controlled.

This angular momentum mixing in the color LOFF order is quite
unique to the dynamically screened magnetic interaction. On writing
the highest eigenvalue of the kernel, Eq.~(\ref{eqvuron}), with angular
momentum $J$, $E_J=E_J^{(0)} +O(\ln^{-1}{1\over\epsilon})$, we have
$E_J^{(0)}=1$ for all $J$ here. On the other hand, for a BCS model with
a pointlike interaction, pairing only occurs in the $s$-wave channel,
and $E_J^{(0)}=\delta_{J,0}$.  Even for a Yukawa type
of interaction, say the screened electric interaction alone, $E_J^{(0)}$
would be different for different $J$.  In both the pointlike and Yukawa
cases, the angular momentum only mixes in higher order corrections, and
would not lead to the enhancement seen in the color LOFF state.

\section*{Acknowledgments}
We wish to thank K.~Rajagopal for providing useful insight and comments.
This work is supported in part by the US Department of Energy under grants
DE-FG02-91ER40651-TASKB and DE-FG02-95ER40899-TASKG.

\ifpreprintsty\else
\end{multicols}
\fi

\end{document}